# Chip-Based Lithium-Niobate Frequency Combs

Mengjie Yu, Cheng Wang, *Member, IEEE*, Mian Zhang, *Member, IEEE* and Marko Lončar, *Senior Member, IEEE*

*Abstract*— **Lithium niobate is an excellent photonic material with large $\chi^{(2)}$ and $\chi^{(3)}$ nonlinearities. Recent breakthroughs in nanofabrication technology have enabled ultralow-loss nanophotonic devices based on a lithium-niobate-on-insulator (LNOI) platform. Here we present an overview of recent developments in the LNOI platform for on-chip optical frequency comb generation. These devices could lead to new opportunities for integrated nonlinear photonic devices for classical and quantum photonic applications.**

## I. INTRODUCTION

ON-CHIP nonlinear photonics has been rapidly advancing due to the continued breakthroughs in ultralow-loss integrated device fabrication. In particular, the strong optical confinement in nanoscale waveguides allows for enhanced nonlinearity and powerful dispersion engineering. As a result, phase-locked optical frequency combs (OFC) can now be generated on centimeter-size chips and can be powered by a battery [1]. To date, OFC's have been demonstrated on various chip-based platforms including silica, silicon nitride, silicon, diamond, fluorides, aluminum nitride and aluminum-gallium arsenide [1]. More importantly, the field of on-chip OFC has quickly moved forward to tackle real-world applications including frequency metrology, precision spectroscopy, distance measurement (LIDAR), searching for exoplanets and optical communications [1,2]. While most OFC platforms rely on third-order nonlinearity ($\chi^{(3)}$) for spectral broadening, they often lack a strong second order nonlinearity ($\chi^{(2)}$) due to material restrictions. As a result, most OFC applications to date still require bulky off-chip $\chi^{(2)}$ components for *f-2f* self-referencing as well as amplitude and phase control of individual comb lines. The lack of on-chip $\chi^{(2)}$ nonlinearity has become a major bottleneck for building fully functional integrated photonic modules based on OFC's.

Lithium niobate (LN) is an ideal material to address this bottleneck. As the workhorse of optoelectronic industry for decades, it is most well-known for its large $\chi^{(2)}$ nonlinearity ($r_{33} = 30$ pm/V) [3]. In addition, LN's refractive index ($n \sim 2.2$) and $\chi^{(3)}$ nonlinearity ($n_2 = 2 \times 10^{-19}$ m²/W) are both comparable to silicon nitride, which has been demonstrated to be an excellent platform for comb generation [1]. LN is also spectrally transparent from 0.35 to 5 μm, and doesn't suffer from two-photon absorption at near-infrared wavelengths. However, LN has long been perceived as a very difficult material to etch and thus not compatible for chip-scale integration. Recent advances in the fabrication of photonic components on the lithium-niobate-on-insulator (LNOI) platform have led to waveguides with 3 dB/m propagation loss and micro-resonators with $Q$ factors more than 10 million [4-10], close to state-of-the-art silicon nitride microrings (37 million $Q$ for similar device dimension [11]). The simultaneous presence of $\chi^{(2)}$ and $\chi^{(3)}$ nonlinearities in LN is opening up exciting opportunities for next-generation integrated OFC technologies with higher levels of integration and advanced functionalities on the same chip. In this review, we focus on the recent developments of LNOI-based OFC devices including Kerr frequency combs, resonant electro-optic (EO) combs, and supercontinuum generation (SCG). We discuss the potential of a fully integrated LN chip with advanced functionalities suitable for nonlinear frequency conversion, frequency metrology, optical communication and precision spectroscopy.

## II. KERR FREQUENCY COMB

Given the ultrahigh $Q$ factors recently seen in LNOI microresonators, the generation of Kerr combs is readily achievable if appropriate dispersion properties could be obtained. Wang, *et al.* have first demonstrated LNOI-based Kerr OFC with bandwidths spanning > 700 nm when pumped at the telecommunication wavelengths (Fig. 1) [12]. Anomalous group-velocity dispersion (GVD) was achieved for both transverse electric (TE) and transverse magnetic (TM) polarizations, leveraging the birefringent properties of LN [Fig. 1(b,c)]. Shortly afterwards, researchers from several different groups have achieved soliton modelocking in LNOI waveguides with similar geometries [9,10]. Specifically, He, *et al.* and Gong *et al.* achieved soliton states using TE polarization in Z-cut LN microring resonators, at 1.5 μm and 2.0 μm, respectively [9,10]. We note that in both demonstrations, soliton mode-locking has been achieved for light polarized

Manuscript submitted August 31, 2019. This work was supported by the National Science Foundation (NSF) (ECCS-1740296 E2CDA), Defense Advanced Research Projects Agency (DARPA) (W31P4Q-15-1-0013), and Air Force Office of Scientific Research (AFOSR) (FA9550-19-1-0310).

Mengjie Yu and Marko Lončar are with the John A. Paulson School of Engineering and Applied Sciences, Harvard University, Cambridge, MA 02138, United States (email: mjyu@g.harvard.edu; loncar@seas.harvard.edu).

Cheng Wang is with the Department of Electrical Engineering and the State Key Laboratory of Terahertz and Millimeter Waves, City University of Hong Kong, Kowloon, Hong Kong, China (email: cwang257@cityu.edu.hk).

Mian Zhang is with HyperLight Corporation, 501 Massachusetts Avenue, Cambridge, MA 02139, United States (email: mian@hyperlightcorp.com).

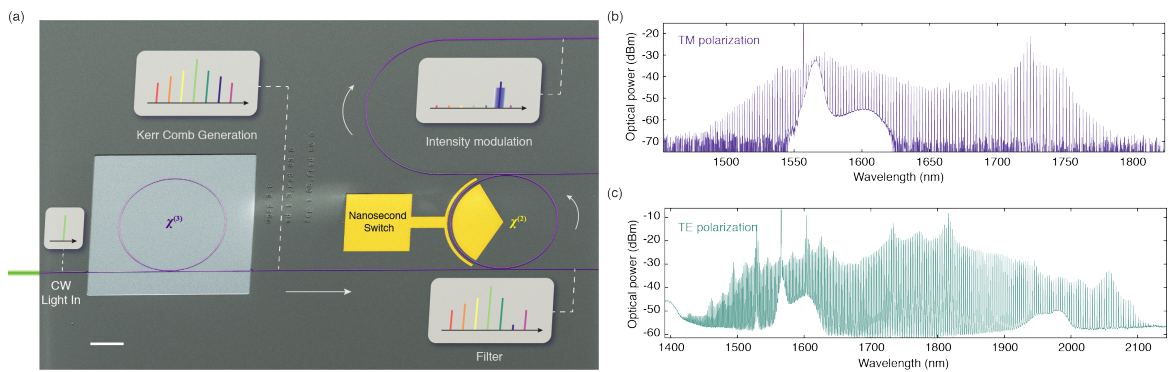

**Fig. 1. Lithium niobate based Kerr frequency combs.** (a) A false-colored scanning electron microscope (SEM) image showing a fabricated LN nanophotonic circuit that consists of a $\chi^{(3)}$-based Kerr comb generator (b) and a $\chi^{(2)}$-based electrically tunable add-drop filter. (b,c) Generated Kerr frequency comb spectra pumping at TM and TE modes. [12] Copyright 2019, Nature Communication **10** (2019).

along the non-polar axis of LN (X and Y axis). This is due to the existence of a strong Raman effect in LN corresponds to the various phonon branches observed in the comb spectra [10, 12-14]. The Raman effect in LN, which is a crystalline material, competes with the four-wave mixing process and can prevent soliton modelocking, especially when the light fields are polarized along the polar axis (Z-axis) [13,14]. However, the realization of combs supporting light polarized along the Z-axis would be ideal for EO manipulation of generated signal as the $\chi^{(2)}$ effect is strongest for that orientation. In order to suppress the Raman oscillation, one could use a smaller LN microresonator with a large FSR (ideally > 500 GHz) [15] or operate at a longer pump wavelength [10].

Besides Kerr comb generation, Wang *et al.* went one step further to show that other non-trivial functionalities available in LNOI photonics could be integrated on the same chip as the comb generator [Fig. 1(a)] [12]. An electrically programmable add-drop filter is monolithically integrated to arbitrarily pick out one specific line from the comb spectrum, and to modulate the intensity of the selected comb line. Although this is only a proof-of-concept demonstration with modulation data rates of < 500 Mbit/s (limited by the photon lifetime inside the resonator), it nevertheless shows that LN-comb-based photonic circuits could play an important role towards fully integrated microcomb applications, including coherent data communication, spectroscopy, and quantum photonics.

The next important milestone for LNOI-based Kerr combs is to achieve an octaves-spanning comb that allows *f-2f* self-referencing. The self-referencing could possibly be simplified by the second harmonic generation (SHG) process intrinsically available in LN [16-18]: the high peak power of the cavity soliton in a modelocked state could contribute to an efficient SHG process, though likely phase-matched to a higher-order transverse mode family, within the same LN microresonator. For example, He *et al.* shows a simultaneous SHG comb generation spanning 200 nm from a modelocked Kerr OFC [9]. With further cavity mode engineering, an octave spanning Kerr OFC combined with simultaneous SHG process would allow for direct self-referencing on the same device. This is particularly appealing for future on-chip optical-clock/frequency-metrology applications since current *f-2f* self-referencing systems require bulky translational laser and periodically-poled LN (PPLN) components [19], which introduces additional noise and complexity to the system.

## III. ELECTRO-OPTIC FREQUENCY COMB

In light of LN's strong $\chi^{(2)}$ nonlinearities, a more interesting question to ask is whether wide spanning OFC's could be generated purely based on $\chi^{(2)}$ processes. In fact, the history of EO ($\chi^{(2)}$) comb generation could be traced back to several decades ago. Conventional EO combs are generated by passing a continuous-wave pump laser through a sequence of amplitude and phase modulators [20-23], which have excellent stability and spectral flatness. However, this scheme results in a limited optical bandwidth and requires high RF power consumption due to the weak EO modulation strength of commercial LN modulators based on titanium-indiffusion or proton-exchange waveguides [24]. The EO interaction strength is dramatically improved in the LNOI platform with wavelength-scale optical confinement [25-29], since metal electrodes can now be placed close to the optical waveguides while maintaining ultralow optical losses. Recently, Wang *et al.* demonstrated LNOI-based EO modulators which can operate at 40 GHz and 100 GHz with half-wave voltages ($V_\pi$) of 1.4 V and 4.4 V [25]. The dramatically improved voltage-bandwidth performances allow for the generation of broader EO combs with high repetition rates. For example, a LNOI-based phase modulator driven at 30 GHz can generate an EO comb spanning > 10 nm [30].

More importantly, much broader EO combs can be realized by embedding the modulation process inside a high-$Q$ optical resonator [31-35] where the harmonics of the free spectral range (FSR) matches the RF driving frequency [Fig. 2(a)]. The light is modulated over multiple roundtrips before it couples out of the cavity, therefore greatly boosting the sideband generation efficiency. Zhang *et al.* demonstrated a resonant EO comb by combining ultra-high $Q$ factor LN cavities (Q > 1,500,000) with fast and low-loss EO modulation [35]. The generated broadband EO comb spectrum spans 80 nm with 900 comb lines spaced by 10 GHz with an RF peak voltage of 10 V. The current system demands microwave power delivered by an off-chip RF amplifier, yet the generated bandwidth is more than 2 orders of magnitude larger than previously reported on-chip EO combs thanks to the ability to engineer the waveguide dispersion. Moreover, the repetition rate and spectral span can be further increased by driving at harmonics of the FSR, as the 30-GHz comb shown in Fig. 2 (d).

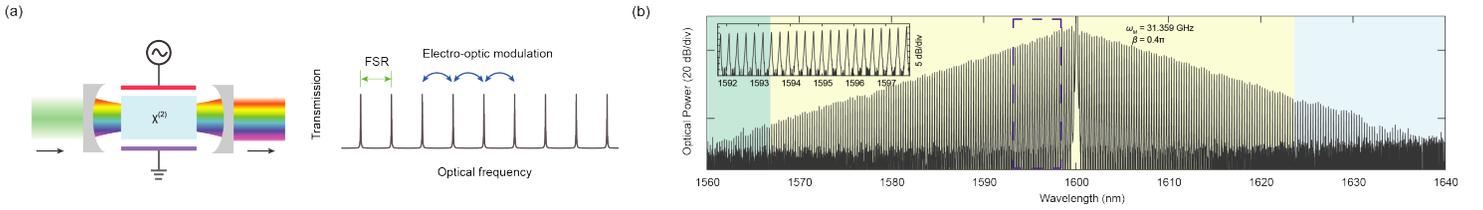

**Fig. 2. Lithium niobate based electro-optic (EO) combs.** (a) Schematic of a resonant EO comb generator. Different cavity modes are coupled via a microwave signal with its modulation frequency equal to the free spectral range (FSR). [35] Copyright 2019, Nature **568** (2019). (b) Generated EO comb driven by a 30-GHz microwave in the 10-GHz-FSR microring.

One key feature of chip-based resonant EO combs is the inherently phase-locked spectra and pulse emission as a result of the coherent low-noise RF drive. In addition, the generation process is not limited by the GVD at the pump wavelength or thermal instability and is highly reconfigurable in operation wavelength and repetition rate [35,36]. Leveraging these properties, dual-comb spectroscopy is recently demonstrated with good mutual coherence and signal-to-noise ratio [36]. One interesting future direction would be to pursue octave-spanning EO combs with optimized GVD along with the assistance of broadband Kerr gain [37]. Meanwhile, an on-chip microwave resonator could significantly reduce the RF power consumption and enable co-integration of photonics and electronic components. We also envision LN-based EO combs opening new opportunities in the visible [7] and mid-infrared (mid-IR) regime where Kerr OFC's face significant challenges in terms of dispersion engineering and material loss.

## IV. SUPERCONTINUUM GENERATION

An alternative way of broadband comb generation is SCG via injecting ultrashort pulses into a nonlinear optical waveguide. Facilitated by the high optical confinement and ultralow propagation loss, integrated photonic waveguides have shown the capability of generating octave-spanning SCG spectra at 10-pJ-level optical pulse energies [38]. Among various integrated photonics platforms, LNOI waveguides are particularly interesting for ultrabroad-band SCG due to the presence of additional $\chi^{(2)}$-based nonlinear interactions such as SHG [17], sum frequency generation (SFG) and different frequency generation (DFG), which naturally converts optical frequencies over large spectral separation.

SCG spanning bandwidths over an octave assisted by SHG have been demonstrated in dispersion-engineered LN waveguides pumping at telecommunication wavelengths [39,40]. Lu *et al.* reported 1.5 octave spanning SCG with 800 pJ of pulse energy for a polarization parallel to the non-polar axis on a Z-cut wafer [39]. Yu *et al.* demonstrated 2.6 octave spanning SCG using 186 pJ of pulse energy in a 5-mm-length LN waveguide for a polarization along the LN polar axis shown in Fig. 3 [40]. Here, the GVD of the LN waveguides are carefully engineered to optimize the spectral overlap between the dispersive wave and the SHG spectrum. The beatnote between the highly coherent SCG and the SHG signals in the regime of spectral overlap allows for direct carrier-envelope-offset frequency $f_{CEO}$ detection with a signal-to-noise ratio of 30 dB [Fig. 3(c)], showing the potential of SCG on LN integrated platforms for future on-chip frequency metrology [40]. Compared with previous demonstration of direct $f_{CEO}$ detection in aluminum nitride (0.8 nJ) [41], LN SCG devices feature significantly lower pulse energy due to its larger $\chi^{(2)}$. Further reduction of the required pulse energy for self-referencing is critical in order to advance the technology beyond laboratory use. Apart from SHG, DFG in LNOI waveguides could lead to broadband mid-IR OFC that is promising for compact spectroscopy sources [42].

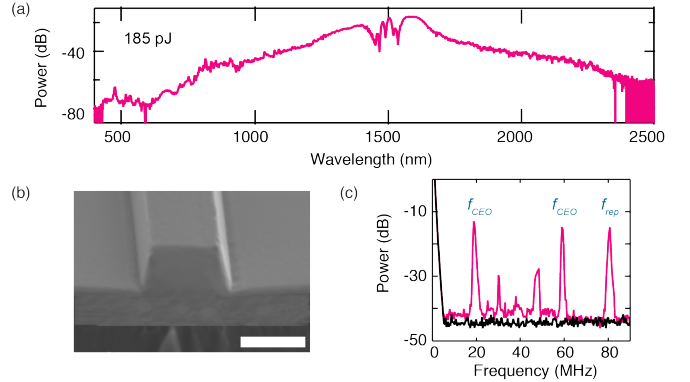

**Fig. 3. Lithium niobate based supercontinuum generation (SCG).** (a) 2.6-octave-spanning coherent supercontinuum spectrum at a pulse energy of 185-pJ. (b) Scanning electron microscope picture of the air-clad LN waveguide cross section. (c) RF spectrum of the comb output. The simultaneous SHG with a coherent SCG allows for the detection of $f_{CEO}$ with a SNR of 30 dB. [40] Copyright 2019, Optics Letters **44** (2019).

Another unique property of LN is that the domain orientation can be inverted when the material is exposed to an external electrical field. A periodical domain inversion of LN (PPLN) can be used to drastically enhance the effective $\chi^{(3)}$ nonlinearity via a cascaded $\chi^{(2)}$ process [43]. Jankowski *et al.* demonstrated more than one octave SCG driven at as low as 2pJ of pulse energy [44], an order of magnitude lower in pulse energy than the conventional PPLN crystals [43]. Dispersion engineering is also employed to enable an ultrabroadband SHG process at 2μm, and a cascade of mixing processes is observed along with SCG. We anticipate significant advances of PPLN technologies [45] as integrated comb sources in the near future.

## V. CHALLENGES AND OUTLOOK

The past few years have witnessed a number of critical breakthroughs in the fabrication techniques and device implementations of the low-loss LNOI photonic platform. These have enabled implementation of LN based systems on the same chip, allowing not only for compact form factors but more importantly orders of magnitude performance improvements. While the results are promising, there are still many challenges and key questions to be answered: 1) How does the photorefractive effect, which induces various kinds of instabilities on different time scales, affect the device integration in thin film LN waveguides? 2) Can octave-spanning modelocked Kerr combs be realized, despite strong Raman effect in crystalline LN material? 3) What are the opportunities offered by a hybrid photonic platform approach that combines Si or Si$_3$N$_4$ platform with the LNOI one? 4) Can the microresonator based EO comb bandwidth be further expanded to reach a full octave, for example using a higher-$Q$ and dispersion engineered optical resonator, and/or a lower-loss efficient microwave interface. 5) What are the opportunities offered by integration of additional LN-based devices with comb generators, including acoustic optical modulators, amplitude and phase modulators, frequency shifters, and waveshapers? What about integration with lasers and detectors on the same chip? We expect a rapid development of LN photonic circuits spanning from visible to the mid-IR regime for OFC and its applications including optical communications, microwave photonics, precision metrology, and spectroscopy.